\documentclass[11pt, a4paper]{article}
\pdfoutput=1

\usepackage[height=8.85in,width=6.75in]{geometry}

\usepackage{graphicx,rotating}     
\usepackage[bookmarksopen,colorlinks=true,linkcolor=light_blue,
citecolor=light_pink,urlcolor=dark_red,linktoc=all]{hyperref}

\usepackage{amsmath,tensor}
\usepackage{mathtools}
\usepackage{amsfonts}
\usepackage{amssymb}
\usepackage{slashed}
\usepackage{braket}
\usepackage{bm}
\usepackage{bbm}
\usepackage{cite}
\usepackage{comment}
\usepackage{mathrsfs}
\usepackage{xcolor}
\usepackage[utf8]{inputenc}
\usepackage[footnotesize]{caption}
\usepackage{bbold}
\usepackage{xfrac}
\usepackage{physics}

\usepackage[lf]{Baskervaldx} 
\usepackage[bigdelims,vvarbb]{newtxmath} 
\usepackage[cal=boondoxo]{mathalfa} 

\usepackage{chngcntr}
\counterwithin{equation}{section}

\usepackage{multicol}
\definecolor{dark_red}{rgb}{0.7, 0., 0.}
\definecolor{light_pink}{rgb}{1,0.4,0.4}
\definecolor{light_blue}{rgb}{0.284602,0.317763,0.963947}
\definecolor{cred}{RGB}{180,50,40}
\definecolor{darkgreen}{RGB}{0, 100, 0}
\definecolor{desy_blue}{HTML}{009EE2}
\definecolor{desy_orange}{HTML}{FD8800}
\definecolor{forestgreen}{HTML}{228B22}
\definecolor{ochre}{HTML}{CCAA2B}

\newcommand{\Mpl}{M_{\text{Pl}}}

\begin{document}

\hypersetup{pageanchor=false}
\begin{titlepage}

\begin{center}

\hfill KEK-TH-2598\\
\hfill RESCEU-3/24\\
\hfill CTPU-PTC-24-05

\vskip .6in

{\Huge \bfseries
Starobinsky Inflation and beyond\\ in Einstein--Cartan Gravity\\
}
\vskip .8in

{\Large Minxi He$^{a}$, Muzi Hong$^{b,c}$, Kyohei Mukaida$^{d,e}$}

\vskip .3in
\begin{tabular}{ll}
$^a$& \!\!\!\!\!\emph{Particle Theory and Cosmology Group, Center for Theoretical Physics of the Universe, }\\[-.3em]
& \!\!\!\!\!\emph{Institute for Basic Science (IBS),  Daejeon, 34126, Korea}\\
$^b$& \!\!\!\!\!\emph{Research Center for the Early Universe (RESCEU), Graduate School of Science, }\\[-.3em]
& \!\!\!\!\!\emph{University of Tokyo, Tokyo 113-0033, Japan}\\
$^c$& \!\!\!\!\!\emph{Department of Physics, Graduate School of Science, }\\[-.3em]
& \!\!\!\!\!\emph{University of Tokyo, Tokyo 113-0033, Japan}\\
$^d$& \!\!\!\!\!\emph{Theory Center, IPNS, KEK, 1-1 Oho, Tsukuba, Ibaraki 305-0801, Japan}\\
$^e$& \!\!\!\!\!\emph{Graduate University for Advanced Studies (Sokendai), }\\[-.3em]
& \!\!\!\!\!\emph{1-1 Oho, Tsukuba, Ibaraki 305-0801, Japan}
\end{tabular}

\end{center}
\vskip .6in

\begin{abstract}
\noindent
We show that various types of scalaron-induced inflation, including the Starobinsky inflation, can be realized in the Einstein--Cartan gravity with the Nieh--Yan term and/or the Holst term.
Einstein--Cartan $f(R)$ theory is known not to induce an additional scalar degree of freedom, the scalaron, contrary to the case in the metric formalism.
However, there exist geometric quantities other than the Ricci scalar in the Einstein--Cartan gravity, such as the Nieh--Yan and the Holst terms.
Once we introduce them in addition to the Ricci scalar and allow general combinations up to their quadratic order, the scalaron can become dynamical to realize inflation.
With the rank of the associate matrix of the quadratic part to be one, the models are equivalent to the $\alpha$-attractor inflation and its deformation, including the Starobinsky inflation and quadratic chaotic inflation, etc.
For more general cases with the rank greater than one, the models fall into the $k$-essence, realizing the rank one case in a particular limit.
\end{abstract}

\end{titlepage}

\tableofcontents
\thispagestyle{empty}
\renewcommand{\thepage}{\arabic{page}}
\renewcommand{\thefootnote}{$\natural$\arabic{footnote}}
\setcounter{footnote}{0}
\newpage
\hypersetup{pageanchor=true}

\section{Introduction}
\label{sec:intro}

Cosmic inflation~\cite{Starobinsky:1980te,Sato:1980yn,Guth:1980zm,Mukhanov:1981xt,Linde:1981mu,Albrecht:1982wi}\footnote{
  See Ref.~\cite{Sato:2015dga} for a review.
}
is the principal paradigm that not only solves the major obstruction of Big Bang cosmology, \textit{i.e.}, the flatness and horizon problems, but predicts the primordial fluctuations consistent with the observations of the cosmic microwave background (CMB).
Since proposed, numerous inflationary models have been put forward and examined, while the observational bounds have become more and more stringent to exclude many well-motivated models.
Currently, the tightest bound on the $(n_s, r)$-plane is imposed by Planck and BICEP/Keck, which in particular yields $r_{0.05} < 0.036$ at $95$\% confidence~\cite{BICEP:2021xfz}.
The Starobinsky inflation~\cite{Starobinsky:1980te} stands out among remaining viable inflationary models since the prediction excellently agrees with the current observation~\cite{Planck:2018jri,BICEP:2021xfz} despite its simplicity.
The only thing necessary for a successful inflation is to add the squared Ricci curvature, $\alpha R^2$, on top of the Einstein--Hilbert action.
It is known that a large coefficient is required, $\alpha \simeq 5\times 10^8 $ for $54$ $e$-folds~\cite{Starobinsky:1983zz,Faulkner:2006ub}, to match the amplitude of CMB fluctuations.
With the help of the Legendre transform, one may readily confirm that this theory involves a dynamical scalar field~\cite{Barrow:1988xi,Barrow:1988kf,Maeda:1987xf,Maeda:1988ab}, \textit{i.e.,} scalaron, which serves as inflaton.
In Einstein's General Relativity (GR), the fundamental building block is the metric that measures the distances, and the field strength of gravity is solely determined by the metric, \textit{i.e.,} the affine connection is assumed to be the Levi-Civita one, which is sometimes referred to as the \textit{metric formalism}.
The Starobinsky inflation mentioned above is based on this metric formalism.
However, from the genuine geometric viewpoint, there is no a priori reason to take the Levi-Civita connection.
In general, the metric and the affine connection can be fully independent.
The most general form of such extensions is the so-called \textit{metric-affine gravity} (see Ref.~\cite{Blagojevic:2012bc} and references therein), while there also exist several restricted classes.

One of the well-known restricted theories is the \textit{Palatini formalism}~\cite{Ferraris:1982wci}, where the affine connection is metric-independent but does not involve the torsion.
Interestingly, in Palatini formalism, it is known that the addition of the squared Ricci curvature never leads to the scalaron~\cite{Sotiriou:2006qn,Sotiriou:2008rp,Olmo:2011uz}, and hence the Starobinsky inflation is not obtained.
The essential reason for this failure is basically because the Ricci curvature does not transform non-trivially under the Weyl transformation, which is originated from the metric-independence of the affine connection.
As anticipated by this observation, the absence of scalaron holds for the $f(R)$ theory in general once we take the affine connection to be metric-independent.
Indeed, as shown in Sec.~\ref{sec:absence}, the scalaron is absent even in the $f(R)$ theory in \textit{Einstein--Cartan (E--C) gravity}, where the affine connection is metric-independent and -compatible but allows the torsion.

The main purpose of this paper is to provide a successful scalaron-induced inflation, such as the Starobinsky inflation, in the E--C gravity.
In the E--C gravity, there exists geometric quantities other than the Ricci curvature, such as the Nieh--Yan (N--Y) term~\cite{Nieh:1981ww} and the Holst term~\cite{Hojman:1980kv,Nelson:1980ph,Castellani:1991et,Holst:1995pc}, owing to the presence of the torsion.
We show that the scalaron becomes dynamical by allowing the N--Y and/or the Holst terms together with the Ricci scalar to be present with general combinations up to their quadratic order.
We demonstrate that this class of models can lead to successful inflation consistent with the current CMB observations, such as the Starobinsky inflation, its $\alpha$-attractor deformation~\cite{Kallosh:2013yoa}, the further deformation other than the $\alpha$-attractor, for example.

\paragraph{Organization of this paper.}
In Sec.~\ref{sec:fR}, after briefly summarizing the minimal basics of the E--C gravity in Sec.~\ref{sec:EC_primer}, we show that the dynamical scalaron does not exist in $f(R)$ theories of the E--C gravity in Sec.~\ref{sec:absence}.
In Sec.~\ref{sec:staro}, we introduce the N--Y term and/or the Holst term together with the Ricci scalar, and consider general combinations up to the quadratic order.
We first restrict ourselves to the case with the addition of the N--Y term \ref{sec:model} and then add the Holst term later in Sec.~\ref{sec:Holst}.
In both cases, the successful scalaron-induced inflation is demonstrated, which includes the Starobinsky inflation as an example.
Sec.~\ref{sec:sum} is devoted to conclusions and discussion.

\section{Absence of scalaron in Einstein--Cartan $f(R)$ gravity}
\label{sec:fR}

To set the stage, we will first introduce the relevant basics of the E--C gravity as well as its relation with the metric and the Palatini formalisms in this section.
We choose the $ f(R) $ theory as an illustrating example to explain why such theories contain a new scalar degree of freedom besides the two graviton modes in the metric formalism but no new dynamical degrees of freedom in the Palatini or E--C case.
We adopt the convention for Riemann tensor and torsion shown in Appendix~\ref{app-convention} which is the same as those in Ref.~\cite{He:2023vlj}.

\subsection{Einstein--Cartan gravity primer}
\label{sec:EC_primer}

The E--C formalism is a subset of the most general metric-affine formalism of gravity where the affine connection $ \bar{\Gamma}\indices{^\rho_{\mu\nu}} $ that describes the geodesic property is treated \textit{a priori} independent of the spacetime metric $ g_{\mu \nu} $.
Generally, this means that the affine connection can be different from the Levi-Civita connection~\eqref{eq-levi-civita-connection} adopted in GR, \textit{i.e.},
\begin{align}\label{eq-general-connection}
  \bar{\Gamma}\indices{^\rho_{\mu\nu}} = \Gamma\indices{^\rho_{\mu\nu}} + C\indices{^\rho_{\mu\nu}} ~,
\end{align}
where the distortion tensor $ C\indices{^\rho_{\mu\nu}} $ denotes the deviation from the Levi-Civita connection $ \Gamma\indices{^\rho_{\mu\nu}} $ (see \textit{e.g.,} \cite{Shimada:2018lnm}).
The distortion tensor is determined by the Euler-Lagrange equation of $ \bar{\Gamma} $.
In the E--C gravity, it is assumed that the affine connection is compatible with metric but the lower indices in $ \bar{\Gamma}\indices{^\rho_{\mu\nu}} $ are not symmetric, which naturally leads to the existence of torsion
\begin{equation}\label{eq-torsion}
  T\indices{^\rho_{\mu\nu}} \equiv \bar{\Gamma}\indices{^\rho_{\mu\nu}} - \bar{\Gamma}\indices{^\rho_{\nu\mu}} =C\indices{^\rho_{\mu\nu}} -C\indices{^\rho_{\nu\mu}} = - T\indices{^\rho_{\nu\mu}} ~.
\end{equation}
As a result, the affine connection deviates by a contorsion tensor from the Levi-Civita connection
\begin{align}
  \bar{\Gamma}\indices{^\rho_{\mu\nu}} &= \Gamma\indices{^\rho_{\mu\nu}} + \frac{1}{2} \left( T\indices{^\rho_{\mu\nu}} + T\indices{_\nu^\rho_\mu} - T\indices{_{\mu\nu}^\rho} \right)  ~.
\end{align}
In the metric formalism, torsion is further chosen to be zero so the affine connection is nothing but $ \Gamma \indices{^\rho_{\mu\nu}} $.
It is known that the Ricci scalar can be rewritten in terms of the metric part and the torsion part as~\cite{Shapiro:2014kma}
\begin{equation}\label{eq-decomposed-R}
  \bar{R} (\bar \Gamma, g) \equiv g^{\mu\nu}\bar{R}_{\mu\nu}(\bar{\Gamma}) = R + 2 \nabla_{\mu} T^{\mu} - \frac{2}{3} T_{\mu} T^{\mu} + \frac{1}{24} S_{\mu} S^{\mu} + \frac{1}{2} q^{\mu\nu\rho} q_{\mu\nu\rho} ~,
\end{equation}
where $ R \equiv \bar{R} (\Gamma) $ is uniquely determined by $ g_{\mu\nu} $ and we have used the decomposition of torsion into the three independent components, \textit{i.e.}, vector $ T_\mu $, axial vector $ S^\mu $, and tensor $ q_{\mu\nu\rho} $ (see Appendix~\ref{app-convention} for details).
The covariant derivative $ \nabla $ on the right hand side of Eq.~\eqref{eq-decomposed-R} is associated with the Levi-Civita connection of $ g_{\mu\nu} $.
Throughout this paper, $ q_{\alpha\beta\gamma} $ plays no significant role because it is always quadratic $ \propto q_{\alpha\beta\gamma} q^{\alpha\beta\gamma} $
by itself.
Solving the constraint equation for $ q_{\alpha\beta\gamma} $ only ends up with trivial solution.
So we will omit it from now on for notation brevity.

Besides, in the E--C gravity (or more general metric-affine gravity), some additional terms are often included as an extension to GR~\cite{Nieh:2008btw,Shapiro:2014kma}.
In this paper, we restrict ourselves to the N--Y term~\cite{Nieh:1981ww} and the Holst term~\cite{Holst:1995pc} which are solely determined by torsion. These terms are recently considered in the context of Higgs inflation with non-minimal coupling with Higgs field~\cite{Langvik:2020nrs,Shaposhnikov:2020gts,He:2023vlj}.
The N--Y term is expressed in terms of the axial vector part of $ T\indices{^\rho_{\mu\nu}} $ as
\begin{equation}\label{eq-NY}
  \int \dd^4x ~ \partial_{\mu} \left( \sqrt{-g} E^{\mu\nu\rho\sigma} T_{\nu\rho\sigma} \right) = -\int \dd^4x ~ \partial_{\mu} \left( \sqrt{-g} S^\mu \right)
  =
  \int \sqrt{- g} \dd^4 x ~ \nabla_\mu S^\mu ~,
\end{equation}
while the Holst term can be decomposed as
\begin{equation}\label{eq-Holst}
  \int \dd^4x ~ \sqrt{-g} E^{\mu\nu\rho\sigma} \bar{R}_{\mu\nu\rho\sigma} = \int \sqrt{- g} \dd^4 x ~ \left( \nabla_\mu S^\mu -\frac{2}{3} T_{\mu} S^{\mu} + \frac{1}{2} E^{\rho\sigma\mu\nu} q_{\lambda \rho \sigma} q\indices{^\lambda_{\mu\nu}} \right) ~,
\end{equation}
from which one can see that the first term on the right hand side is actually the N--Y term.
The N--Y term is obviously a total derivative, \textit{i.e.}, a topological term, so simply adding it to the action does not affect the equation of motion.
On the other hand, the Holst term cannot be expressed as a total derivative like the N--Y term, but it can be easily shown, by varying the action with respect to $ T_{\mu} $, $ S^{\mu} $, $ q_{\alpha\beta\gamma} $, that torsion is constrained to be trivial in the case where the action only contains the Einstein--Hilbert term and the Holst term.
Therefore, the addition of the N--Y term and the Holst term does not change the dynamics of the system.
Again, the $ q_{\mu\nu\rho} $ contribution is trivial throughout this paper so we will omit it from the Holst term for notation simplicity.
In Sec.~\ref{sec:staro}, however, the N--Y term and the Holst term will play an essential role in our main discussion.
Noticing that Eq.~\eqref{eq-Holst} is a combination of the N--Y term and $ S_{\mu}T^{\mu} $ (up to a $ q_{\mu\nu\rho} $-related term), the contribution from the Holst term can be converted into that from $ S_{\mu}T^{\mu} $.
Therefore, we will treat $ \nabla_{\mu} S^{\mu} $ and $ S_{\mu}T^{\mu} $ as two fundamental ingredients in the discussion of effects of the N--Y and the Holst terms for convenience.

\subsection{Absence of scalaron}
\label{sec:absence}

Now we are ready to discuss the $ f (R) $ gravity as a simple example and illustrate the relation among metric, Palatini, and E--C formalisms.
Given the action
\begin{align}\label{eq-action-fR}
  S= \int \sqrt{-g_{\rm J}} \dd^4 x\, f(\bar{R}(\bar{\Gamma}, g_{\rm J})) ~,
\end{align}
where we assume that $ f''(\bar{R}) \neq 0 $ (otherwise it trivially coincides with GR) and ``J'' denotes Jordan frame, one can perform the Legendre transformation to extract the information from the functional $ f(\bar{R}) $ and to simplify the situation, by introducing an auxiliary field $ \chi $, such that
\begin{align}\label{eq-fR-Legendre}
  S= \int \sqrt{-g_{\rm J}} \dd^4 x\, \left[ \left( \bar{R}(\bar{\Gamma}, g_{\rm J}) -\chi \right) f'(\chi) +f(\chi) \right] ~.
\end{align}
The equation of motion for $ \chi $ can be easily derived and solved, which shows that $ \chi $ dynamical coincides with $ \bar{R} $.
Substitution of the solution for $ \chi $ recovers the original action~\eqref{eq-action-fR}.
Inserting Eq.~\eqref{eq-general-connection} and solving the equation of motion for $ C\indices{^\rho_{\mu\nu}} $, one finds
\begin{align}\label{eq-fR-C-solution}
  C\indices{^\rho_{\mu\nu}} = \frac{1}{2} \left( \delta_\mu^\rho \partial_\nu \ln \frac{2f'}{\Mpl^2} +\delta_\nu^\rho \partial_\mu \ln \frac{2f'}{\Mpl^2}- g_{\rm J}^{\rho\sigma} g_{{\rm J}\mu\nu} \partial_{\sigma} \ln \frac{2f'}{\Mpl^2} \right) +\delta_\nu^\rho U_\mu ~,
\end{align}
where $ U_\mu $ is an arbitrary vector that cannot be determined by the equation of motion.
In order to fix the connection, one can impose additional conditions such as metricity for E--C gravity or torsionlessness for Palatini formalism.
In the former case, $ U_\mu $ can be chosen as $ U_\mu = -\frac{1}{2} \partial_\mu \ln (2f'/\Mpl^2) $ to have a torsionful connection that is compatible with spacetime metric.
In the latter case, one can simply set $ U_\mu =0 $ such that $ C\indices{^\rho_{\mu\nu}} $ is symmetric with respect to the lower two indices and, therefore, torsionless.
In other words, one can freely choose $ U_\mu $ to adopt either E--C or Palatini formalism.

This arbitrary vector affects the properties of the affine connection, but Ricci scalar $ \bar{R} $ remains invariant under a shift of connection by a vector because of the projective symmetry which can be explicitly proved by the definition of Ricci scalar, \textit{i.e.}, $ \bar{R} $ is unchanged with
\begin{align}\label{eq-projective-transformation}
  g_{\mu\nu} \to g_{\mu\nu} ~, \qquad
  \bar{\Gamma}\indices{^\rho_{\mu\nu}} \to \bar{\Gamma}\indices{^\rho_{\mu\nu}} +\delta_{\nu}^{\rho} U_{\mu} ~.
\end{align}
Thus, E--C and Palatini formalisms are equivalent up to an arbitrary vector or a gauge choice in $ f(\bar{R}) $.
Substituting the solution~\eqref{eq-fR-C-solution} back to the action leads to\footnote{
  In Refs.~\cite{Inagaki:2022blm,Kumar:2024bfc}, they overlooked a constraint equation when redefining $f'$ with a scalar field, leading to the wrong conclusion. This is basically because they have not introduced $(\bar R - \chi) f' (\chi)$ that guarantees the equivalence after transformation. Once we take into account the constraint equation, we arrive at the same action as ours and hence results in the absence of scalaron.
}
\begin{align}
  S= \int \sqrt{-g_{\rm J}} \dd^4 x\, \left[ f'(\chi) \left( R_{\rm J} + \frac{3}{2} g_{\rm J}^{\mu\nu} \partial_\mu \ln \frac{2f'}{\Mpl^2} \partial_\nu \ln \frac{2f'}{\Mpl^2} \right) +f(\chi) -\chi f'(\chi) \right] ~.
\end{align}
The contribution from $ C\indices{^\rho_{\mu\nu}} $ apparently serves as a kinetic term for $ \ln (2f'/\Mpl^2) $ that could have been identified as a new scalar degree of freedom.
However, this is not true.
One easy way to see that is to transform into the Einstein frame\footnote{Extracting the conformal mode~\cite{Ema:2020zvg} and analyzing the kinetic structure~\cite{He:2023vlj} can also show that the superficial kinetic term for $ \ln (2f'/\Mpl^2) $ does not lead to a new degree of freedom. See Appendix~\ref{app-conformal} for detail.}.
Performing a conformal transformation to the above action,
\begin{align}\label{eq-conformal-transf-fR}
  g_{{\rm E}\mu\nu} = \frac{2f'}{\Mpl^2} g_{{\rm J}\mu\nu} ~,
\end{align}
where ``E'' denotes the Einstein frame, the Jordan frame Ricci scalar is transformed as follows
\begin{align}\label{eq-R-conformal-transformation}
  R_{\rm J} = \frac{2f'}{\Mpl^2} R_{\rm E} -6 \left( \frac{2f'}{\Mpl^2} \right)^{3/2} \Box_{\rm E} \left( \frac{2f'}{\Mpl^2} \right)^{-1/2} ~,
\end{align}
where $ \Box \equiv g^{\mu\nu} \nabla_\mu \nabla_\nu $ represents the d'Alembert operator.
The second term in Eq.~\eqref{eq-R-conformal-transformation} will finally cancel the contribution from $ C\indices{^\rho_{\mu\nu}} $ such that there will be no new degree of freedom.
As a result, $ f(R) $ theory in E--C or Palatini formalism does not contain a new scalar degree of freedom\footnote{This is true even in general metric-affine gravity, which can be easily seen by applying conformal transformation~\eqref{eq-conformal-transf-fR} directly to Eq.~\eqref{eq-fR-Legendre} where $ \bar R $ transforms trivially.}
On the other hand, $ C\indices{^\rho_{\mu\nu}} $ is set to be zero from the beginning, so the second term in Eq.~\eqref{eq-R-conformal-transformation} then becomes the kinetic term without cancellation, which leads to the appearance of a new scalar degree of freedom, \textit{i.e.}, the scalaron~\cite{Starobinsky:1980te}\footnote{$ f(R) $ theory in Palatini formalism has been studied intensively studied in Refs.~\cite{Sotiriou:2006qn,Sotiriou:2008rp,Olmo:2011uz}, and our results of the absence of scalaron are consistent with them.
The review of metric case can be found, for instance, in Ref.~\cite{DeFelice:2010aj}.}.

The above discussion explicitly explains the reason why, in the E--C $ f(R) $ gravity or especially $ R+R^2 $ case, we do not expect a new degree of freedom other than two graviton modes to play a role, for example, as the inflaton like the Starobinsky model~\cite{Starobinsky:1980te}.
However, in the rest of this paper, we will show that this situation can be changed once we include the N--Y and/or the Holst terms.

\section{Dynamical scalaron in Einstein--Cartan gravity}
\label{sec:staro}

In this section, we show that the dynamical scalaron exists in several models of Einstein--Cartan gravity that include the N--Y term or the combination of the N--Y term and the Holst term.
We consider models with operators up to dimension four with $\bar{R}$ and these two terms and classify them into two categories.

In Sec.~\ref{sec:model}, we restrict our discussion to models consisting of only $\bar R$ and the N--Y term.
The general model in this category contains fewer terms than those including the Holst term additionally, and it is possible to study it extensively.
We declare the general model and show two specific examples that lead to quadratic chaotic inflation~\cite{Linde:1983gd} and $\alpha$-attractor inflation~\cite{Kallosh:2013yoa} in Sec.~\ref{sec:NY_quadratic} and Sec.~\ref{sec:NY_Starobinsky}, respectively.
The general model and its limits are analyzed in Sec.~\ref{sec:NY_general}.

As detailed below, the analysis of Sec.~\ref{sec:NY_general} shows that among models with only the N--Y term, those with an associate matrix of rank one lead to theories with a canonical kinetic term and a potential, while others with an associate matrix of a higher rank lead to the $k$-essence, or equivalently $p(\phi,X)$, theory.
In Sec.~\ref{sec:Holst}, we investigate models in the other category which include both the N--Y term and the Holst term.
We mainly focus on the models with rank one, which may describe slow roll inflation.
A brief discussion of cases with higher rank is also presented for completion.

\subsection{Models with Nieh--Yan term}
\label{sec:model}

The general model including the N--Y term with operators up to dimension four with only $\bar R$ and the N--Y term is
\begin{equation}
  S = \int \sqrt{-g} \dd^4 x\, \qty[
    \frac{\Mpl^2}{2} \bar R
    + \alpha_\text{R} \bar R^2
    + \alpha_\text{RS} \bar R\, \nabla_\mu S^\mu
    + \alpha_\text{S} \qty( \nabla_\mu S^\mu )^2
  ]~,
  \label{eq-NY_general}
\end{equation}
where we drop the arguments of Ricci scalar for notational brevity.
Note that the last three terms in the square brackets in Eq.~(\ref{eq-NY_general}) are of quadratic form.
We rewrite Eq.~(\ref{eq-NY_general}) as
\begin{equation}
  S = \int \sqrt{-g} \dd^4 x\, \qty[
    \frac{\Mpl^2}{2} \bar R
    + {\bm O}^\mathsf{t} \, \mathbb{S} \, {\bm O}
  ]~,
  \label{eq-NY_general_matrix}
\end{equation}
with the vector ${\bm O}^\mathsf{t} = (\bar R, \nabla_\mu S^\mu)$ and the symmetric matrix
\begin{equation}
  \mathbb{S}  =
  \begin{pmatrix}
    \alpha_\text{R} & \frac{1}{2}\alpha_\text{RS} \\
    \frac{1}{2}\alpha_\text{RS} & \alpha_\text{S} \\
    \end{pmatrix}~.
  \label{eq-matrixA}
\end{equation}
For non-zero $\alpha_\text{R}$, it is convenient for later use to change the variables as ${\bm O}_\text{D} = \mathbb{P}^{-1} {\bm O}$ with
\begin{equation}
  \mathbb{P}^{-1} =
  \begin{pmatrix}
    1 & \frac{\alpha_\text{RS}}{2 \alpha_\text{R}} \\
    0 & 1 \\
    \end{pmatrix}~,
\end{equation}
which gives the diagonalized form:
\begin{equation}
  {\bm O}^\mathsf{t}\, \mathbb{S} \, {\bm O}
  =
  {\bm O}^\mathsf{t}_\text{D}  \qty(\mathbb{P}^\mathsf{t}\, \mathbb{S}\, \mathbb{P}) {\bm O}_\text{D}
  =
  \alpha_\text{R} \qty( \bar R + \frac{\alpha}{2} \nabla_\mu S^\mu )^2
  + \beta \qty( \nabla_\mu S^\mu )^2 ~,
  \label{eq-standard_form_NY}
\end{equation}
where $\alpha \equiv \alpha_\text{RS} / \alpha_\text{R}$ and $\beta \equiv \alpha_\text{S} - \alpha_\text{RS}^2 / (4\alpha_\text{R})$.
According to Sylvester's law of inertia, for a general real symmetric matrix $\mathbb{S}$ of order $n$ and a non-singular matrix $\mathbb{P}$ which makes $\mathbb{D} = \mathbb{P}^\mathsf{t}\, \mathbb{S}\, \mathbb{P}$ a diagonal matrix, the number of positive, negative, and null elements in the diagonal of $\mathbb{D}$ is always the same regardless of the choice of $\mathbb{P}$.
Hence, the rank of $\mathbb{S}$ is equivalent to the number of nonzero terms in the diagonalized form ${\bm O}^\mathsf{t}_\text{D} \, \mathbb{D} \, {\bm O}_\text{D}$ of the quadratic form.

In the following, we consider the Legendre transform of ${\bm O}$ to extract the dynamical scalar mode in Eq.~\eqref{eq-NY_general_matrix}.
The Legendre transform is defined for a convex function, which in our case implies that the matrix $\mathbb{S}$ given in \eqref{eq-matrixA} fulfills $\det \mathbb{S} \neq 0 $, \textit{i.e.}, $\alpha_\text{R} \alpha_\text{S} \neq \alpha_\text{RS}^2/4$.
In  this case, one may rewrite the quadratic form by means of dual variables ${\bm \gamma}$ as
\begin{equation}
  {\bm O}_\text{D}^\mathsf{t}\, \mathbb{D} \, {\bm O}_\text{D}
  - \qty({\bm \chi} - \mathbb{D}\, {\bm O}_\text{D})^\mathsf{t}\, \mathbb{D}^{-1} \qty( {\bm \chi} - \mathbb{D}\, {\bm O}_\text{D}) =
  - {\bm \chi}^\mathsf{t}\, \mathbb{D}^{-1} {\bm \chi} + 2\, {\bm \chi} \cdot {\bm O}_\text{D} ~.
\end{equation}
Note that the second term in the left-hand side can be understood as an insertion of unity as it gives a trivial Gaussian path integral.
In the limit of one element of $\mathbb{D}$ to be zero, the corresponding component of ${\bm \chi}$ is forced to be zero, which is consistent with $\det \mathbb{D} = 0$, \textit{i.e.}, $\alpha_\text{R} \alpha_\text{S} = \alpha_\text{RS}^2/4$.
Under the assumption that $\mathbb{D}$ is not a null matrix, the matrix given in \eqref{eq-matrixA} has rank one in this case, and the Legendre transform is performed only for a vector with nonzero element of $\mathbb{D}$.
There are three possibilities for rank one.
$\alpha_\text{S} = \alpha_\text{RS} = 0$ leads to the discussion in Sec.~\ref{sec:absence}.
We will discuss the case of $\alpha_\text{R} = \alpha_\text{RS} = 0$ in Sec.~\ref{sec:NY_quadratic} and the case of $\alpha_\text{R},\,\,\alpha_\text{RS} \neq 0$ while $\beta \equiv \alpha_\text{S} - \alpha_\text{RS}^2 / (4\alpha_\text{R}) = 0$ in Sec.~\ref{sec:NY_Starobinsky}.
The case of rank two will be discussed in Sec.~\ref{sec:NY_general}.

\subsubsection{Quadratic chaotic inflation}
\label{sec:NY_quadratic}
Let us start our discussion with the simplest example of inflation in our setup, where we take $\alpha_\text{R} = \alpha_\text{RS} = 0$.
Although the resultant inflation model is already excluded by observations, we believe its demonstration is instructive because of the simplicity.

The starting point is
\begin{equation}
  S = \int \sqrt{-g} \dd^4 x\, \qty[
    \frac{\Mpl^2}{2} \qty(R - \frac{2}{3} T_{\mu} T^{\mu} + \frac{1}{24} S_{\mu} S^{\mu})
    + \beta \qty( \nabla_\mu S^\mu )^2
  ]~,
  \label{eq-quardratic1}
\end{equation}
where we have used the decomposition of $ \bar R $~\eqref{eq-decomposed-R} and dropped the total derivative terms.
Hereafter, we will always do the decomposition without mentioning again.
As discussed in the previous section, we introduce an auxiliary field $\chi$ as follows:
\begin{equation}
  S = \int \sqrt{-g} \dd^4 x\, \qty[
    \frac{\Mpl^2}{2} \qty(R - \frac{2}{3} T_{\mu} T^{\mu} + \frac{1}{24} S_{\mu} S^{\mu})
    + 2 \chi \nabla_\mu S^\mu - \frac{\chi^2}{\beta}
  ]~.
\end{equation}
By solving the constraint equation of $\chi$, one can recover (\ref{eq-quardratic1}).
After integrating by parts and solving constraint equations of $T^\mu$ and $S^\mu$ respectively, one obtains
\begin{equation}
  S = \int \sqrt{-g} \dd^4 x\, \qty(
    \frac{\Mpl^2}{2} R
    - \frac{48}{\Mpl^2} \partial_\mu \chi \partial^\mu \chi
    - \frac{\chi^2}{\beta}
  )~.
\end{equation}
Defining $\sigma = \sqrt{96} \chi / \Mpl$, we rewrite the action as
\begin{equation}
  S = \int \sqrt{-g} \dd^4 x\, \qty(    \frac{\Mpl^2}{2} R
    - \frac{1}{2} \partial_\mu \sigma \partial^\mu \sigma
    - \frac{\Mpl^2}{96 \beta} \sigma^2
  )~,
  \label{eq-NY_quadratic_final}
\end{equation}
which characterizes the quadratic chaotic inflation with $\beta > 0$.

\subsubsection{Starobinsky inflation and its deformation}
\label{sec:NY_Starobinsky}

Now we consider the case of $\beta = 0$ while $\alpha_\text{R}, \alpha \neq 0$.
As we will see shortly, the action coincides with the Starobinsky inflation for $\alpha = 1$ and the coupling $\alpha$ serves as a mass deformation parameter in the same way as the $\alpha$-attractor~\cite{Kallosh:2013yoa}.

Here we also introduce an auxiliary field and obtain
\begin{align}\label{eq-NY-Star-deform}
  \begin{split}
  S_\text{J} = \int \sqrt{-g_\text{J}} \dd^4 x\, & \left[
    \frac{\Mpl^2}{2} \qty(R_\text{J} - \frac{2}{3} T_{\mu} T^{\mu} + \frac{1}{24} S_{\mu} S^{\mu})
    \right.\\
    & \left.
    + 2 \chi \qty( R_\text{J} + 2 \nabla_\mu T^\mu - \frac{2}{3} T_{\mu} T^{\mu} + \frac{1}{24} S_{\mu} S^{\mu} + \frac{\alpha}{2} \nabla_\mu S^\mu ) - \frac{\chi^2}{\alpha_\text{R}}
    \right]~.
  \end{split}
\end{align}
The index J means that we are currently at Jordan frame.
As in the last section, one integrates by part and solves constraint equations of $T^\mu$ and $S^\mu$ respectively.
By defining the conformal factor as $\Omega^2 = 1 + 4 \chi / \Mpl^2$, one may rewrite the action as follows
\begin{equation}
  S_\text{J} = \int \sqrt{-g_\text{J}} \dd^4 x\, \qty[
    \frac{\Mpl^2 \Omega^2}{2} R_\text{J}
    + \frac{3}{4} \Mpl^2 \Omega^2 ( \partial_\mu \!\ln \Omega^2)^2
    - \frac{3}{4} \alpha^2 \Mpl^2 \Omega^2 ( \partial_\mu \!\ln \Omega^2)^2
    - \frac{\Mpl^4 (\Omega^2 - 1)^2}{16 \alpha_\text{R}}
  ]~.
\end{equation}
Here we have implicitly assumed that the conformal factor should be strictly positive, $\Omega^2 > 0$, to avoid catastrophic instabilities.
Now we perform the Weyl transformation
\begin{equation}
g_{\text{E} \mu \nu} = \Omega^2 g_{\text{J} \mu \nu}~,
\end{equation}
and go to the Einstein frame. The action is transformed to
\begin{equation}
  S_\text{E} = \int \sqrt{-g_\text{E}} \dd^4 x\, \qty[
    \frac{\Mpl^2}{2} R_\text{E}
    - \frac{3}{4} \alpha^2 \Mpl^2 ( \partial_\mu \!\ln \Omega^2)^2
    - \frac{\Mpl^4 (\Omega^2 - 1)^2}{16 \alpha_\text{R} \Omega^4}
  ]~.
\end{equation}
One may further rewrite the action by means of a canonically normalized field $\sigma = \sqrt{3/2} \alpha \Mpl \ln(\Omega^2)$ as
\begin{equation}
  S_\text{E} = \int \sqrt{-g_\text{E}} \dd^4 x\, \qty[
    \frac{\Mpl^2}{2} R_\text{E}
    - \frac{1}{2} \partial_\mu \sigma \partial^\mu \sigma
    - \frac{\Mpl^4}{16 \alpha_\text{R}} \qty(1 - e^{- \sqrt{\frac{2}{3}} \frac{\sigma}{ \alpha \Mpl}})^{2}
  ]~.
  \label{eq-Starobinsky}
\end{equation}
The resultant model is nothing but the Starobinsky inflation deformed \textit{a la} the $\alpha$-attractor~\cite{Kallosh:2013yoa}, which reproduces the Starobinsky inflation for $\alpha = 1$\footnote{The simplest case $ \alpha =1 $ was first pointed out in Ref.~\cite{Bombacigno:2019nua} and general $ \alpha $ in Ref.~\cite{Boudet:2023phd} in the context of modified gravity.}.
The appearance of a dynamical scalaron in this way can also be expected if one considers the quantum corrected Higgs inflation in E--C gravity~\cite{He:2023vlj} where there are two dynamical inflaton fields, the radial mode of Higgs and the scalaron.
By taking the limit of infinitely large Higgs mass, one expects that the Higgs decouples while the scalaron remains as the only inflaton as shown above.
The existence of a dynamical scalaron can be shown in a frame-independent manner with the help of the conformal mode, which can be seen in Appendix~\ref{app-conformal}.

The scalar spectral index $n_s$ and the tensor-to-scalar ratio $r$ at CMB scale predicted by \eqref{eq-Starobinsky} with $\alpha_\text{R} > 0$ can be expressed, in leading order of large $ N $, as~\cite{Kallosh:2013yoa}
\begin{equation}\label{eq-alpha-attractor-large-N}
  n_s = 1 - \frac{2}{N} ~, \qquad r = \frac{12 \alpha^2}{N^2}~,
\end{equation}
respectively, where the $e$-folding number at the CMB scale is denoted by $N$.
The amplitude of scalar perturbations can be matched to the CMB observation with an appropriate choice of $\alpha_\text{R}$.
The remaining free parameter $\alpha$ allows deformation of the model including such as the quadratic chaotic inflation and the Starobinsky inflation as is well known.

\subsubsection{$k$-essence as general form}
\label{sec:NY_general}

Now we are ready to analyze the full-rank case declared as (\ref{eq-standard_form_NY}).
We introduce two auxiliary fields since there are two independent terms in \eqref{eq-standard_form_NY}.
As we will see shortly, one of them becomes the dynamical scalaron while the other provides a constraint equation.
By solving the constraint equation, we show that the general form of this model is the $k$-essence, \textit{i.e.,} $p(\phi,X)$ theory with $X = - (\partial \phi)^2$.
As a sanity check, we recover the cases in Sec.~\ref{sec:NY_quadratic} and Sec.~\ref{sec:NY_Starobinsky} by taking the limits of parameters appropriately.
The limits of taking one or two parameter(s) in \eqref{eq-NY_general} to be zero are discussed in Appendix~\ref{app-other}.

Let us rewrite \eqref{eq-standard_form_NY} by utilizing two auxiliary fields $(\chi, \chi_\beta)$:
\begin{equation}
  \begin{split}
  S_\text{J} = \int \sqrt{-g_\text{J}} \dd^4 x\, &\left[
    \frac{\Mpl^2}{2} \qty(R_\text{J} - \frac{2}{3} T_{\mu} T^{\mu} + \frac{1}{24} S_{\mu} S^{\mu})
    \right. \\
    & + 2 \chi \qty(R_\text{J} + 2 \nabla_\mu T^\mu  - \frac{2}{3} T_{\mu} T^{\mu} + \frac{1}{24} S_{\mu} S^{\mu} + \frac{\alpha}{2} \nabla_\mu S^\mu ) - \frac{\chi^2}{\alpha_\text{R}} \\
    &\left. + 2 \chi_\beta \nabla_\mu S^\mu - \frac{\chi_\beta^2}{\beta} \right]~.
  \end{split}
  \label{eq-NY_general3}
\end{equation}
By solving the constraint equations of $\chi$ and $\chi_\beta$, respectively, one may recover \eqref{eq-standard_form_NY}.
Now as in the previous sections, from \eqref{eq-NY_general3} one may immediately solve the constraint equations of $T^\mu$ and $S^\mu$, respectively, to obtain
\begin{equation}
  \begin{split}
  S_\text{J} = \int \sqrt{-g_\text{J}} \dd^4 x\, &\left[
    \frac{\Mpl^2}{2} \Omega^2 R_\text{J}
    + \frac{3}{4} \Mpl^2 \Omega^2 (\partial_\mu \ln \Omega^2)^2
    -\frac{12}{\Mpl^2 \Omega^2} \qty(\frac{\alpha}{4} \Mpl^2 \partial_\mu \Omega^2 + 2 \partial_\mu \chi_\beta )^2
    \right. \\
    &\left.
    - \frac{\Mpl^4}{16 \alpha_\text{R}} (\Omega^2 - 1)^2 - \frac{\chi_\beta^2}{\beta}
    \right]~,
  \end{split}
  \label{eq-NY_general4}
\end{equation}
where the conformal factor is $\Omega^2 = 1 + 4 \chi / \Mpl^2$.

For later convenience, we define $\Sigma$ as
\begin{equation}
\sqrt{\frac{2}{3}} \frac{\Sigma}{\Mpl \alpha} \equiv \Omega^2 - 1 + \frac{8 \chi_\beta}{\alpha \Mpl^2}~.
\end{equation}
By further performing the Weyl transformation,
we rewrite the action \eqref{eq-NY_general4} as follows
\begin{equation}
  S_\text{E} = \int \sqrt{-g_\text{E}} \dd^4 x\, \left[
    \frac{\Mpl^2}{2} R_\text{E}
    - \frac{1}{2 \Omega^4} \qty(\partial_\mu \Sigma)^2
    - \frac{\Mpl^4}{16 \alpha_\text{R}} \frac{(\Omega^2 - 1)^2}{\Omega^4}
    - \frac{\alpha^2 \Mpl^4}{64 \beta} \qty(\sqrt{\frac{2}{3}} \frac{\Sigma}{\Mpl \alpha \Omega^2} + \frac{1}{\Omega^2} - 1)^2
    \right]~.
  \label{eq-pX}
\end{equation}
Solving the constraint equation for $\Omega^2$, one finds
\begin{equation}
  \Omega^2 =
  \frac{\partial_\mu \Sigma \partial^\mu \Sigma + \frac{\Mpl^4}{8 \alpha_\text{R}} + \frac{\alpha^2 \Mpl^4}{32 \beta} \qty(\sqrt{\frac{2}{3}} \frac{\Sigma}{\Mpl \alpha} + 1)^2}
  {\frac{\Mpl^4}{8 \alpha_\text{R}} + \frac{\alpha^2 \Mpl^4}{32 \beta} \qty(\sqrt{\frac{2}{3}} \frac{\Sigma}{\Mpl \alpha} + 1)}
  ~.
  \label{eq-Omega_constraint}
\end{equation}
Note here that the positivity of $\Omega^2$ implicitly restricts the allowed range of $\Sigma$.
When $\alpha_\text{R}$, $\alpha$ and $\beta$ all take non-zero values, one ends up with a $p(\Sigma, X)$ theory with $X = - (\partial \Sigma)^2$ as in \eqref{eq-pX}.

We check the consistency by taking the limits that reproduce the previous results for rank one, \textit{i.e.,}  $\alpha_\text{R} \alpha_\text{S} = \alpha_\text{RS}^2/4$, discussed in Secs.~\ref{sec:NY_quadratic} and Sec.~\ref{sec:NY_Starobinsky}.
In the limit of $\alpha_\text{R} \to 0$ with fixed $\alpha$, equivalently $\alpha_\text{R}, \alpha_\text{RS} \to 0$, one immediately obtains
\begin{equation}
  \Omega^2 \to 1~,
  \label{eq-alphaR_limit}
\end{equation}
and reproduces \eqref{eq-NY_quadratic_final} in Sec.~\ref{sec:NY_quadratic}.
On the other hand, for $\beta \to 0$, it is equivalent to
\begin{equation}
  \Omega^2 \to \sqrt{\frac{2}{3}} \frac{\Sigma}{\Mpl \alpha} + 1~.
\end{equation}
Recalling the positivity of $\Omega^2 > 0$, one may define $\sigma$ as $\Omega^2 = \sqrt{\frac{2}{3}} \frac{\Sigma}{\Mpl \alpha} + 1 \equiv \exp \sqrt{\frac{2}{3}} \frac{\sigma}{\alpha \Mpl}$, and reproduces \eqref{eq-Starobinsky} in Sec.~\ref{sec:NY_Starobinsky}.
One may also wonder what kind of theory one can have when taking some of the parameters in \eqref{eq-NY_general} to be zero.
We examined all the cases and it turned out that except for the rank-one models discussed in Sec.~\ref{sec:absence}, Sec.~\ref{sec:NY_quadratic} and Sec.~\ref{sec:NY_Starobinsky}, one ends up with $p(\Sigma, X)$ theory.
See Appendix~\ref{app-other} for the calculations of taking these limits.
The limits can also be recovered if one starts from \eqref{eq-NY_general} with one or two parameter(s) set to be zero and uses auxiliary fields to analyze.

\subsection{Scalaron in the presence of Holst term}
\label{sec:Holst}
In this section, we consider models including also the Holst term\footnote{The case with only the Holst term is considered in Ref.~\cite{Salvio:2022suk}.}, which effectively introduces the term $S_\mu T^\mu$ as explained in Sec.~\ref{sec:EC_primer}.
With a slight generalization to the previous Sec.~\ref{sec:model}, the action can be written as
\begin{equation}
  S = \int \sqrt{-g} \dd^4 x\, \qty[ \frac{\Mpl^2}{2} \qty(\bar R + \zeta S_\mu T^\mu) + {\bm O}^\mathsf{t}\, \mathbb{S}\, {\bm O} ]~,
\end{equation}
where the symmetric matrix is
\begin{equation}
  \mathbb{S} = \begin{pmatrix}
    \alpha_\text{R} & \frac{\alpha_\text{RS}}{2} & \frac{\alpha_\text{RST}}{2} \\
    \frac{\alpha_\text{RS}}{2} & \alpha_\text{S} & \frac{\alpha_\text{SST}}{2} \\
    \frac{\alpha_\text{RST}}{2} & \frac{\alpha_\text{SST}}{2} & \alpha_\text{ST}
  \end{pmatrix}~,
  \label{eq-direct}
\end{equation}
and the vector is ${\bm O} = ( \bar R, \nabla_\mu S^\mu, S_\mu T^\mu )$.
Note that the Holst term allows a nontrivial contribution already at a linear order, \textit{i.e.,} $\zeta S_\mu T^\mu$.
By changing the variables as ${\bm O}_\text{D} = \mathbb{P}^{-1} {\bm O}$ with
\begin{equation}
  \mathbb{P}^{-1} = \begin{pmatrix}
    1 & \frac{\alpha_\text{RS}}{2 \alpha_\text{R}} & \frac{\alpha_\text{RST}}{2 \alpha_\text{R}} \\
    0 & 1 & \frac{\beta_\text{SST}}{2 \beta_\text{S}} \\
    0 & 0 & 1
  \end{pmatrix}~,
\end{equation}
one may diagonalize the quadratic form as follows:
\begin{equation}
  {\bm O}^\mathsf{t}\, \mathbb{S} \, {\bm O}
  =
  {\bm O}^\mathsf{t}_\text{D}  \qty(\mathbb{P}^\mathsf{t}\, \mathbb{S}\, \mathbb{P}) {\bm O}_\text{D}
  = \alpha_\text{R} \qty( \bar R + \frac{\alpha}{2} \nabla_\mu S^\mu + \frac{\tilde\alpha}{2} S_\mu T^\mu )^2 + \beta_\text{S} \qty( \nabla_\mu S^\mu + \frac{\beta}{2} S_\mu T^\mu )^2
  + \gamma_\text{ST} \qty( S_\mu T^\mu )^2 ~,
  \label{eq-OSOHolst}
\end{equation}
where
\begin{equation}
  \begin{split}
    &\alpha \equiv \frac{\alpha_\text{RS}}{\alpha_\text{R}}~, \qquad
    \tilde \alpha \equiv \frac{\alpha_\text{RST}}{\alpha_\text{R}}~,
    \qquad
    \beta_\text{S} \equiv \alpha_\text{S} - \frac{\alpha_\text{R}\alpha^2 }{4}~,
    \qquad
    \beta_\text{SST} \equiv \alpha_\text{SST} - \frac{\alpha_\text{R} \alpha \tilde \alpha}{2}~,
    \qquad
    \beta \equiv \frac{\beta_\text{SST}}{\beta_\text{S}}~, \\[.5em]
    &\beta_\text{ST} \equiv \alpha_\text{ST} - \frac{\alpha_\text{R} \tilde \alpha^2}{4}~,
    \qquad
    \gamma_\text{ST} \equiv \beta_\text{ST} - \frac{\beta_\text{S} \beta^2}{4}~.
  \end{split}
\end{equation}
Note that one assumed $\alpha_\text{R} \neq 0$ and $\beta_\text{S} \neq 0$ in $\mathbb{P} ^ {-1}$, while it is possible to take $\alpha_\text{R} \to 0$ or $\beta_\text{S} \to 0$ limits in (\ref{eq-OSOHolst}).

As in the previous Sec.~\ref{sec:model}, the properties of the model are expected to depend on the rank of $\mathbb{S}$.
The simplest class of models is obtained for $\rank{\mathbb{S}} = 1$ because the non-trivial dependence of $X$ \textit{a la} $p(\phi,X)$ theory does not arise.
In the case of $\alpha_R \neq 0$, the condition for $\rank{\mathbb{S}} = 1$ is given by $0 = \beta_\text{S} = \gamma_\text{ST}$.
Note that from the definition of $\beta$, requiring $\beta_\text{S} = 0$ implies $\beta_\text{SST} = 0$.
The conditions are thus
\begin{equation}
  \alpha_\text{S} = \frac{\alpha_\text{RS}^2}{4 \alpha_\text{R}}~,
  \qquad
  \alpha_\text{SST} = \frac{\alpha_\text{RS}\alpha_\text{RST}}{2 \alpha_\text{R}} ~,
  \qquad
  \alpha_\text{ST} = \frac{\alpha_\text{RST}^2}{4 \alpha_\text{R}}~.
  \label{eq-rank1-NYH}
\end{equation}
In the case of $\alpha_\text{R} = 0$, from the limit of \eqref{eq-rank1-NYH} one observes that
\begin{equation}
  \alpha_\text{S}\alpha_\text{ST} = \frac{\alpha_\text{SST}^2}{4}
  \label{eq-rank1-aR0}
\end{equation}
must hold.
Also, at least one of the parameters in \eqref{eq-rank1-aR0} is not zero.
Thus, generally, the conditions for $\rank{\mathbb{S}} = 1$ are given by
\begin{equation}
  \alpha_\text{R} \alpha_\text{S} = \frac{\alpha_\text{RS}^2}{4}~,
  \qquad
  \alpha_\text{R}\alpha_\text{SST} = \frac{\alpha_\text{RS}\alpha_\text{RST}}{2} ~,
  \qquad
  \alpha_\text{R} \alpha_\text{ST} = \frac{\alpha_\text{RST}^2}{4}~,
  \qquad
  \alpha_\text{S}\alpha_\text{ST} = \frac{\alpha_\text{SST}^2}{4}~,
\end{equation}
where at least one parameter is not zero.
Note that there is one redundant condition if all parameters are non-zero, while all of these are required taking into account all cases.
One can also derive these conditions directly from (\ref{eq-direct}).
If these conditions are met, we need only one auxiliary field,
which will be discussed in detail in Sec.~\ref{sec:NYH_starobinsky}.
Otherwise, for $\rank{\mathbb{S}}=2,3$, the model in general leads to $k$-essence theories as will be briefly discussed in Sec.~\ref{sec:kessence_NYH}.

\subsubsection{Further deformation of Starobinsky inflation}
\label{sec:NYH_starobinsky}

In this section, we consider the case of $\rank{\mathbb{S}} = 1$, \textit{i.e.,} the conditions given in Eq.~\eqref{eq-rank1-NYH} are fulfilled.
We only need one auxiliary field $\chi$ to perform the Legendre transform in this case.
As we will see shortly, by solving the constraint equation for $S_\mu$ and $T_\mu$, we obtain the ``standard'' kinetic term for the auxiliary field, and hence it does not involve the $k$-essence theories.

We start with the most general $\rank{\mathbb{S}} = 1$ model with the Holst term:
\begin{equation}
  \begin{split}
    S = \int \sqrt{-g_\text{J}} \dd^4 x\, &\left[
    \frac{\Mpl^2}{2} \qty(R_\text{J} - \frac{2}{3} T_{\mu} T^{\mu} + \frac{1}{24} S_{\mu} S^{\mu} + \zeta S_\mu T^\mu) \right. \\
    &\left.
    + \alpha_{\text R} \qty(R_\text{J} + 2 \nabla_\mu T^\mu - \frac{2}{3} T_{\mu} T^{\mu} + \frac{1}{24} S_{\mu} S^{\mu} + \frac{\alpha}{2} \nabla_\mu S^\mu + \frac{\tilde \alpha}{2} S_\mu T^\mu )^2
  \right] ~.
  \end{split}
  \label{eq-generalHolst2}
\end{equation}
At this stage, we assume $\alpha_{\text R} \neq 0$. The $\alpha_{\text R} \to 0$ limit will be discussed later.
The action can be expressed by means of an auxiliary field $\chi$ as
\begin{equation}
  \begin{split}
    S = \int \sqrt{-g_\text{J}} \dd^4 x &\left[
      \frac{\Mpl^2}{2} \qty(R_\text{J} - \frac{2}{3} T_{\mu} T^{\mu} + \frac{1}{24} S_{\mu} S^{\mu} + \zeta S_\mu T^\mu) \right. \\
      &\left.
      + 2 \chi \qty(R_\text{J} + 2 \nabla_\mu T^\mu - \frac{2}{3} T_{\mu} T^{\mu} + \frac{1}{24} S_{\mu} S^{\mu} + \frac{\alpha}{2} \nabla_\mu S^\mu + \frac{\tilde \alpha}{2} S_\mu T^\mu )
      - \frac{\chi^2}{\alpha_\text{R}}
      \right] ~.
  \end{split}
  \label{eq-generalHolst-aux}
\end{equation}
Performing the Weyl transformation with the conformal factor of $\Omega^2 \equiv 1 + 4 \chi / \Mpl^2$, we obtain Starobinsky-like inflation with a deformed kinetic term for $ \sigma \equiv \ln \Omega^2 $:
\begin{equation}
  S = \int \sqrt{-g_\text{E}} \dd^4 x\, \qty[
    \frac{\Mpl^2}{2} R_\text{E}
    - \frac{\Mpl^2}{2} \frac{3 \qty{\alpha e^\sigma + 3 \qty[ \zeta + \qty( e^\sigma - 1 ) \tilde \alpha/2 ]}^2}{2 e^{2 \sigma} + 18 \qty[ \zeta + \qty( e^\sigma - 1 ) \tilde \alpha /2 ]^2 } \partial_\mu \sigma \partial^\mu \sigma
    - \frac{\Mpl^4}{16 \alpha_{\text R}} \qty(1 - e^{-\sigma})^2
  ]~.
  \label{eq-generalHolst}
\end{equation}

Let us first check the consistency.
For this purpose, we clarify that there exists a redundancy in the description in the presence of the Holst term
in $\zeta = \tilde\alpha/2$.
In this case, one may always eliminate the cross-term of $S_\mu T^\mu$ in Eq.~\eqref{eq-generalHolst-aux} by the following redefinitions of
\begin{equation}
  T_\mu' = T_\mu - \frac{3}{4} \zeta S_\mu~,
  \qquad
  S_\mu' = \sqrt{1 + 9 \zeta^2} S_\mu~,
\end{equation}
while this change can be absorbed by
\begin{equation}
  \alpha' = \frac{\alpha + 3 \zeta}{\sqrt{1 + 9 \zeta^2}}~.
\end{equation}
Hence, in the limit of $\zeta = \tilde\alpha/2$, one obtains the Starobinsky inflation deformed \textit{a la} the $\alpha$-attractor given in Sec.~\ref{sec:NY_Starobinsky} as
\begin{equation}
  S = \int \sqrt{-g_\text{E}} \dd^4 x\, \qty[
    \frac{\Mpl^2}{2} R_\text{E}
    - \frac{1}{2} \partial_\mu \sigma \partial^\mu \sigma
    - \frac{\Mpl^4}{16 \alpha_{\text R}} \qty(1 - e^{-\sqrt{\frac{2}{3}}\frac{\sigma}{\alpha' \Mpl}})^2
  ]~.
  \label{eq-Holst_Starobinsky}
\end{equation}
This result can also be obtained by taking $\zeta = \tilde\alpha/2$ with a suitable field redefinition in Eq.~\eqref{eq-generalHolst}.
If we further take $\alpha' = 0$, \textit{i.e.,} $\alpha = - 3 \zeta = - 3 \tilde\alpha/2$, we should recover the E--C $f(\bar R)$ gravity in Sec.~\ref{sec:absence}.
One may immediately confirm that the kinetic term of $\sigma$ vanishes in Eq.~\eqref{eq-generalHolst}, which agrees with the absence of scalaron in the E--C $f(\bar R)$.
The limit involving $\alpha_\text{R} = 0$ should be treated with care.
As can be inferred from Eq.~\eqref{eq-rank1-NYH}, one may keep $\alpha_\text{S}$ finite but send $\alpha_\text{RS} = 0$, which implies $\alpha_\text{R}\alpha^2 \to 4 \alpha_\text{S}$ at $\alpha_\text{R} \to 0$, \textit{i.e.,} the quadratic chaotic inflation in Sec.~\ref{sec:NY_quadratic}.
Indeed, we can readily show this after an appropriate field redefinition of $\sigma$ in Eq.~\eqref{eq-generalHolst}.

Now it is clear that the general $\rank{\mathbb{S}} = 1$ model for $\zeta \neq \tilde \alpha/2$ can be regarded as deformation of the $\alpha$-attractor Starobinsky inflation.
To illustrate the effect of this modification, we simplify the situation by reducing the number of parameters.
In particular, we consider the limit of $\alpha_\text{R} = 0$ under $\alpha_\text{S}$, $\alpha_\text{ST} \neq 0$, corresponding to $\alpha \to \infty$ but fixing $\alpha_\text{R}\alpha^2/4 = \alpha_\text{S}$ and $\beta = 2 \tilde\alpha/\alpha$, where the kinetic term can be canonically normalized analytically as we see shortly.
This limit is infinitely far away from the $\alpha$-attractor Starobinsky inflation $\zeta = \tilde\alpha/2$ since we take the limit of $\tilde\alpha \to \infty$ but fix $\zeta$ to be finite, and we also expect that the $\alpha$-attractor Starobinsky inflation would be reproduced by taking $\zeta \to \infty$.
The action in this limit reads
\begin{equation}
  S = \int \sqrt{-g_\text{J}} \dd^4 x\, \qty[
    \frac{\Mpl^2}{2}
    \qty( R_\text{J} - \frac{2}{3} T_{\mu} T^{\mu} + \frac{1}{24} S_{\mu} S^{\mu} + \zeta S_\mu T^\mu )
    + \alpha_\text{S} \qty(\nabla_\mu S^\mu + \frac{\beta}{2} S_\mu T^\mu)^2
  ]~.
  \label{eq-SSTaction}
\end{equation}
Either by performing the Legendre transformation of Eq.~\eqref{eq-SSTaction} or by taking the limits directly in Eq.~\eqref{eq-generalHolst}, we find
\begin{equation}
  S = \int \sqrt{-g_\text{E}} \dd^4 x\, \qty[
    \frac{\Mpl^2}{2} R_\text{E}
    - \frac{48 \Mpl^2}{9 (\zeta \Mpl^2 + 2 \beta \chi)^2 + \Mpl^4} \partial_\mu \chi \partial^\mu \chi
    -  \frac{\chi^2 }{\alpha_\text{S}}
  ]~.
  \label{eq-SSTHolst}
\end{equation}

\begin{figure}[t]
  \centering
  \includegraphics[width=0.48\linewidth]{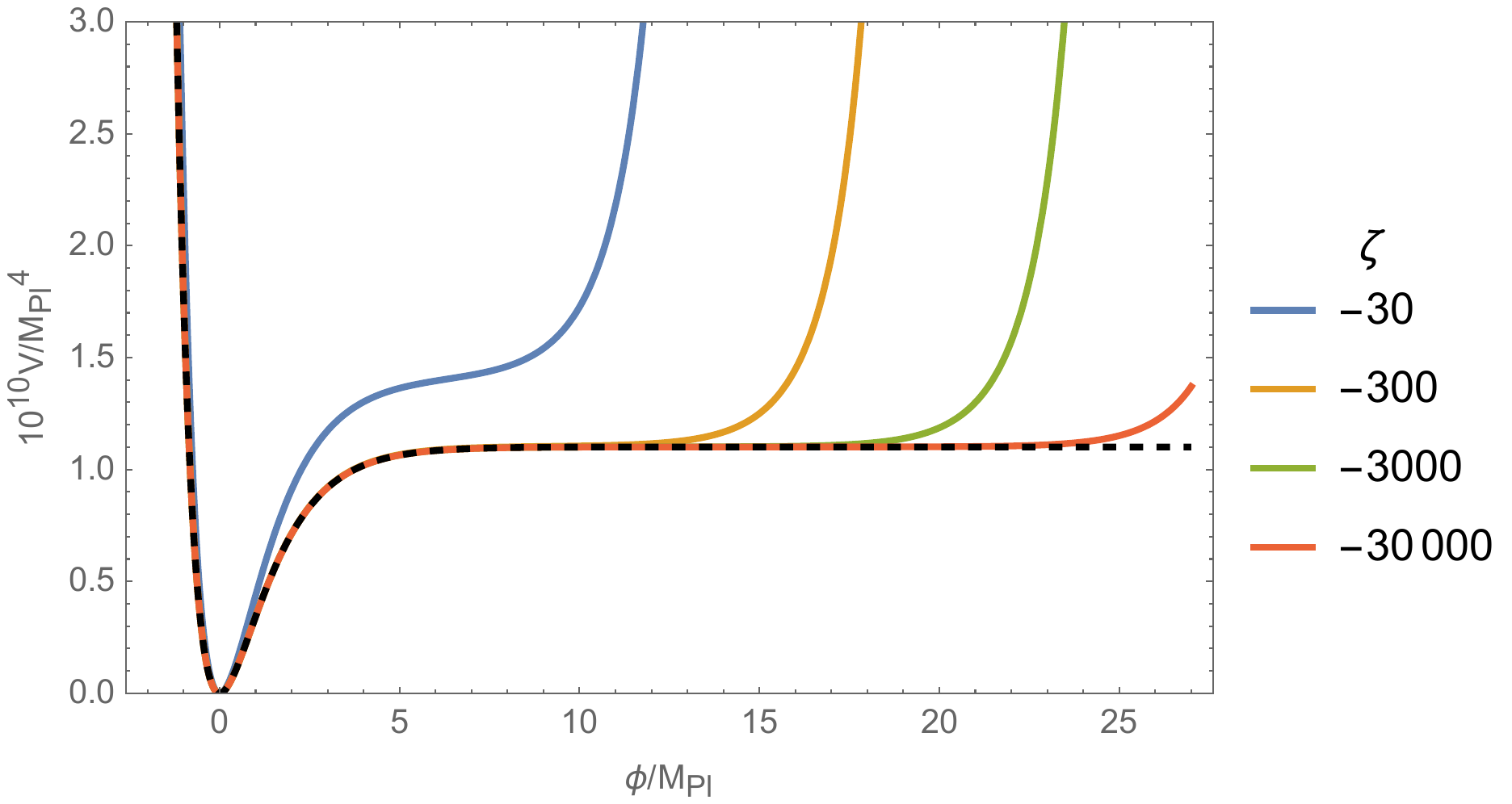}
  \includegraphics[width=0.45\linewidth]{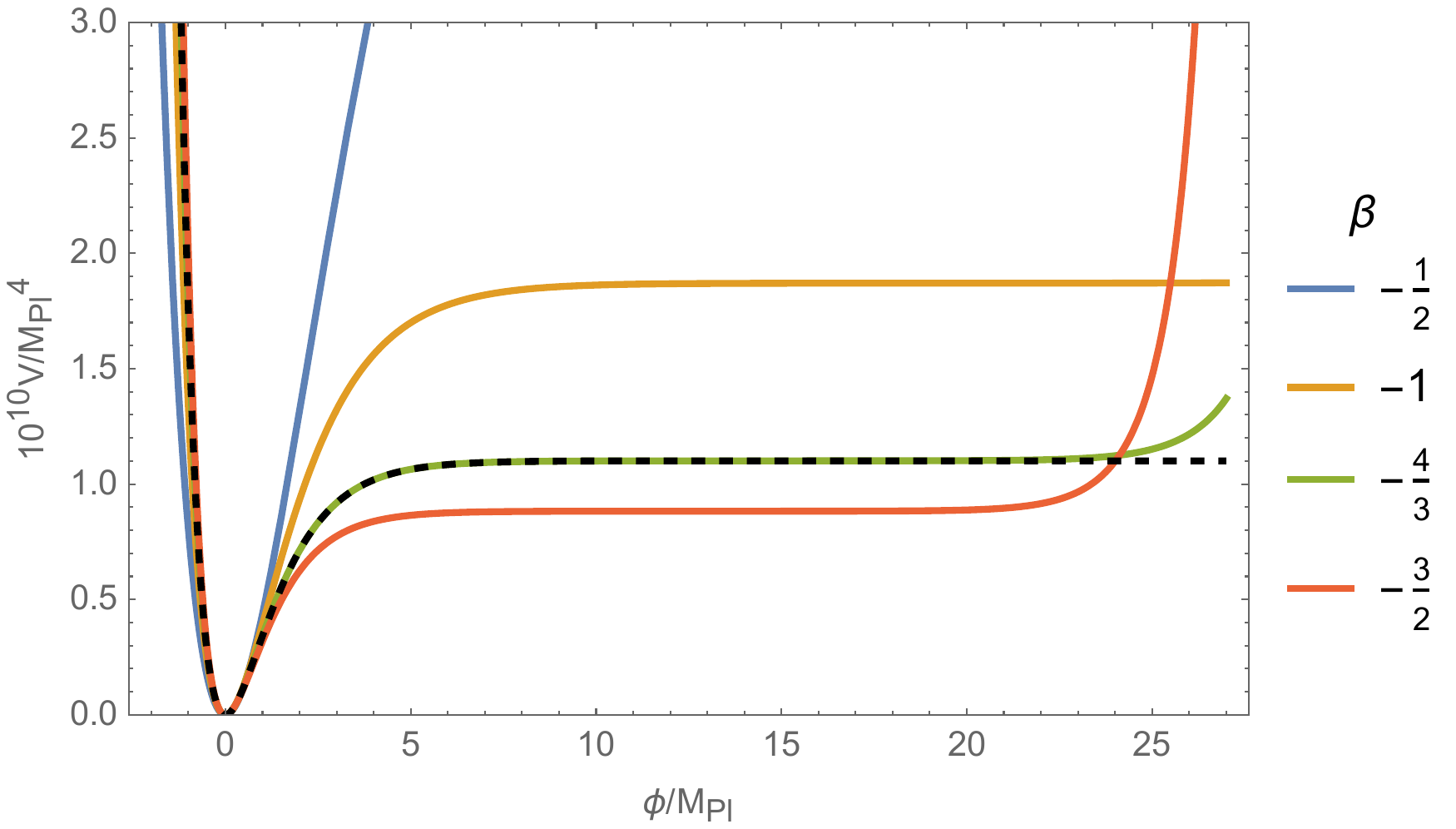}
  \caption{Inflaton potential in Eq.~\eqref{eq-deform-Star-Einstein-frame} with different parameters are shown.
  Black dashed line is the scalaron potential in the Starobinsky model in the Einstein frame, corresponding to $ \beta =-4/3 $ and $ \zeta \to -\infty $.
  Left: varying $ \zeta $ while fixing $ \beta=-4/3 $.
  Right: varying $ \beta $ while fixing $ \zeta= -3\times 10^4 $. }
  \label{fig-potential}
\end{figure}

\begin{figure}[t]
  \centering
  \includegraphics[width=0.8\linewidth]{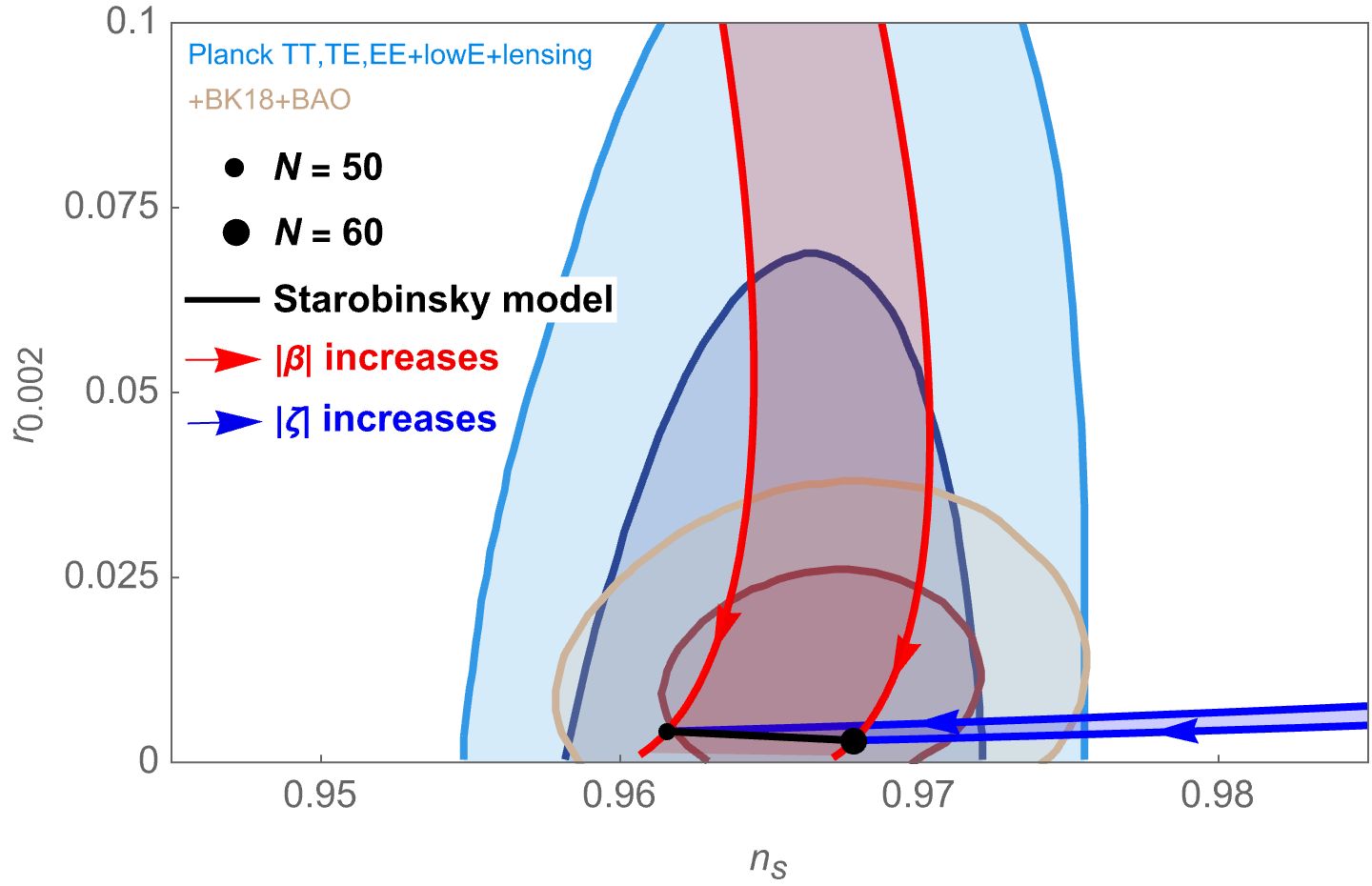}
  \caption{Predictions of spectral index $ n_s $ and tensor-to-scalar ratio $ r $ from the model given in Eq.~\eqref{eq-SSTaction}.
  The constraint contours are directly taken from Fig. 5 in Ref.~\cite{BICEP:2021xfz}.
  The blue contours are observational constraints from Planck~\cite{Planck:2018jri} at pivot scale $ k/a_0=0.002 \,\mathrm{Mpc}^{-1} $ with $1 \sigma$ and $ 2\sigma $ regions respectively.
  The brown contours are constraints combining with BICEP/Keck~\cite{BICEP:2021xfz} at pivot scale $ k/a_0=0.05 \,\mathrm{Mpc}^{-1} $ (and they have assumed the tensor spectral index $ n_t =0 $), with $1 \sigma$ and $ 2\sigma $ regions respectively.
  The Starobinsky model corresponds to $ N=54 $ on the black line.
  The red trajectory (coincides with $ \alpha $-attractor~\cite{Kallosh:2013yoa}) is obtained by fixing $ \zeta = -3\times 10^4 $ while varying $ \beta $ from $ -1/25 $ (corresponding to large $ r $ which lies outside the figure) to $ -5/2 $ (corresponding to small $ r $).
  Note that this trajectory is different from the predictions in Eqs.~\eqref{eq-alpha-attractor-large-N}, namely curved rather than straight, because Eqs.~\eqref{eq-alpha-attractor-large-N} are only at leading order of large $ N $.
  $ \beta =-4/3 $ coincides with the predictions from the Starobinsky model when $ \zeta \to -\infty $ (here $ |\zeta|=3\times 10^4 $ is large enough in numerical calculation).
  The deep blue trajectory is obtained by fixing $ \beta=-4/3 $ while changing $ \zeta $ from $ -13 $ (corresponding to large $ n_s $ which lies outside the figure) to $ -3\times 10^4 $ (corresponding to small $ n_s $).
  Since large-$ |\zeta| $ limit leads to the Starobinsky model, the predictions approach the black line as $ |\zeta| $ increases. }
  \label{fig-inflation-prediction}
\end{figure}

We further transform \eqref{eq-SSTHolst} to have a canonical kinetic term:
\begin{equation}\label{eq-deform-Star-Einstein-frame}
  S = \int \sqrt{-g_\text{E}} \dd^4 x\, \qty[
    \frac{\Mpl^2}{2} R_\text{E}
    - \frac{1}{2} \partial_\mu \phi \partial^\mu \phi
    - \frac{\Mpl^4}{36 \alpha_\text{S} \beta^2}
    \qty( 3 \zeta + \sinh \sqrt{\frac{3}{8}} \frac{\beta \phi}{\Mpl}
    )^2
  ]~.
\end{equation}
The sign of $\zeta$ and $\beta$ can be taken arbitrary as it is absorbed into the redefinition of $\phi \mapsto - \phi$.
Here and hereafter, we take $\zeta < 0$ and $\beta < 0$ without loss of generality.
The shape of the potential is shown in Fig.~\ref{fig-potential} for several sets of parameters.
If the absolute value of $\zeta$ is relatively large, the part of the potential which is relevant to inflation is approximately of Starobinsky-type.
To see this, one can shift the minimum of the potential
\begin{equation}
  \phi_\text{min} =\sqrt{\frac{8}{3}} \frac{\Mpl}{\beta} \ln (- 3 \zeta + \sqrt{1 + 9 \zeta^2}) ~,
\end{equation}
to the origin, which leads to
\begin{equation}
V(\phi)
=
\frac{\Mpl^4}{144 \alpha_\text{S} \beta^2} \qty[ 6 \zeta + e^{\sqrt{\frac{3}{8}} \frac{\beta}{\Mpl} (\phi + \phi_\text{min})} - e^{- \sqrt{\frac{3}{8}} \frac{\beta}{\Mpl} (\phi + \phi_\text{min})}
]^2~.
\end{equation}
If the absolute value of $\zeta$ is large, the second exponential is negligible compared with the first one around the origin, leading to the $\alpha$-attractor Starobinsky inflation.
Hence, the observables, the scalar spectral index $n_s$ and the scalar-to-tensor ratio $r$ at CMB scales is mainly controlled by $\beta$.
For fixed $\beta$, $\zeta$ controls the length of the ``flat part'' of the potential.
The value of $\alpha_\text{S}$ is determined by matching to the scalar fluctuations $\Delta_s$ at CMB scales.
The deviation of the observables of this model from the ones of the $\alpha$-attractor models is shown in Fig.~\ref{fig-inflation-prediction}.

\subsubsection{$k$-essence as general form}
\label{sec:kessence_NYH}

Here we briefly discuss the cases of $\rank{\mathbb{S}} > 1$ and show that they in general lead to $k$-essence.
Since the case of $\rank{\mathbb{S}} = 2$ can be obtained from a certain limit of $\rank{\mathbb{S}} = 3$, we only consider $\rank{\mathbb{S}} = 3$ in the following.
Our starting point is the following action with three auxiliary fields
\begin{equation}
  \begin{split}
    S = \int \sqrt{-g_\text{J}} \dd^4 x &\left[
    \frac{\Mpl^2}{2} \qty(R_\text{J} - \frac{2}{3} T_{\mu} T^{\mu} + \frac{1}{24} S_{\mu} S^{\mu} + \zeta S_\mu T^\mu) \right. \\
    &
    + 2 \chi \qty(R_\text{J} + 2 \nabla_\mu T^\mu - \frac{2}{3} T_{\mu} T^{\mu} + \frac{1}{24} S_{\mu} S^{\mu} + \frac{\alpha}{2} \nabla_\mu S^\mu + \frac{\tilde \alpha}{2} S_\mu T^\mu )
    - \frac{\chi^2}{\alpha_\text{R}} \\
    &
    \left.
    + 2 \chi_\beta \qty( \nabla_\mu S^\mu + \frac{\beta}{2} S_\mu T^\mu ) - \frac{\chi_\beta^2}{\beta_\text{S}}
    + 2 \chi_\gamma S_\mu T^\mu - \frac{\chi_\gamma^2}{\gamma_\text{ST}}
    \right]~.
  \end{split}
\end{equation}
One may readily solve the constraint equations for $T_\mu$ and $S_\mu$ and obtain
\begin{equation}
  \begin{split}
    S = \int \sqrt{-g_\text{E}} \dd^4 x
    &\left\{
      \frac{\Mpl^2}{2} R_\text{E}
      - \frac{3}{4 \Mpl^2 \Omega^4 \qty( 1 + 9 G^2 )} \qty[ \qty( 3 G + \alpha ) \Mpl^2 \partial \Omega^2 + 8 \partial \chi_\beta ]^2
    \right. \\
    &\left.
      - \frac{\Mpl^4}{16\alpha_\text{R}} \qty( 1 - \Omega^{-2} )^2
      - \frac{\chi_\beta^2}{\beta_\text{S} \Omega^4} - \frac{\chi_\gamma^2}{\gamma_\text{ST} \Omega^4}
    \right\}~,
  \end{split}
  \label{eq-k-NYH}
\end{equation}
where the conformal factor is defined as $\Omega^2 \equiv 1 + 4 \chi / \Mpl^2$ and a function in the kinetic term is
\begin{equation}
  G (\Omega^2, \chi_\beta, \chi_\gamma) \equiv
  \Omega^{-2} \qty[ \zeta + \frac{\tilde\alpha}{2} \qty( \Omega^2 - 1 ) + \frac{\beta}{2} \frac{4 \chi_\beta}{\Mpl^2} + \frac{4 \chi_\gamma}{\Mpl^2} ]~.
\end{equation}
Obviously, the scalaron is identified as a particular linear combination of auxiliary fields and the other two auxiliary degrees of freedom are non-dynamical.
Since the kinetic term depends on non-dynamical degrees of freedom in Eq.~\eqref{eq-k-NYH}, we in general end up with an action involving a non-trivial rational function of the scalaron kinetic term, \textit{i.e.,} $k$-essence, by solving the constraint equations for non-dynamical fields.

\section{Conclusions and discussion}
\label{sec:sum}

In this paper, we have studied the E--C gravity up to dimension four operators that consist of the Ricci scalar, the N--Y term and/or the Holst term.
Contrary to the E--C $f(R)$ gravity (Sec.~\ref{sec:absence}), it has been found that the scalaron becomes dynamical in general owing to the presence of the N--Y term and/or the Holst term.
The property of the model is classified by the rank of the symmetric matrix $\mathbb{S}$ that specifies the quadratic form in the variables of the Ricci scalar, the N--Y term and/or the Holst term.
We have shown that the Starobinsky inflation and its mass-deformed variant \textit{a la} $\alpha$-attractor is realized in the rank-one case, in general with the N--Y term (Sec.~\ref{sec:NY_Starobinsky}) and in a certain limit with both the N--Y and the Holst terms (Sec.~\ref{sec:NYH_starobinsky}), which are perfectly consistent with the current CMB observations.
We have discussed how the inflationary prediction on the $(n_s , r)$ plane gets modified by turning on some parameters other than the $\alpha$-attractor Starobinsky limit (Sec.~\ref{sec:NYH_starobinsky}).
For a fully general model parameter with the rank being greater than one, our Lagrangian exhibits the $k$-essence, \textit{i.e.,} $p (X, \phi)$ theory, providing a smooth connection among different models (Secs.~\ref{sec:NY_general} and \ref{sec:kessence_NYH}).

We have only considered the combination of the N--Y term and the Holst term, but in principle one may also consider terms like $ \nabla_\mu T^\mu $, $ T^\mu T_\mu $, the Euler class, and Pontryagin class, etc.
The effect of $ \nabla_\mu T^\mu $ is essentially the same as the N--Y term, \textit{i.e.}, modifying the kinetic term of the scalaron by adding one new parameter dependence, which can also lead to the Starobinsky inflation.
The $ \nabla_\mu T^\mu $ contribution associated with $ \bar R $ in the quadratic part essentially plays a role to cancel the kinetic term for the conformal factor from the conformal transformation, as shown in Sec.~\ref{sec:absence}.
If additional $ \nabla_\mu T^\mu $ is taken into account, the coefficient of it should be constrained within a finite range of negative value such that the new contribution to the kinetic term will not lead to ghost degree of freedom by inducing a kinetic term with a wrong sign.
Furthermore, if $ T^\mu T_\mu $ is also considered, this constraint may be altered.
The Euler class and Pontryagin class are topological invariants as the N--Y term, but the former are already of mass dimension four by themselves, so, considering up to dimension-four operators, they can only appear as they are without combination with other terms or themselves and, thus, their contribution is trivial.
One can find more complicated theories with torsion in Ref.~\cite{Mikura:2023ruz}.

Either the N--Y term or the Holst term involves a $CP$ odd operator, and hence the chiral current coupling of the scalaron to the matter sector, such as chiral fermions in the Standard Model (SM), is predicted~\cite{Shaposhnikov:2020gts,Salvio:2022suk}.
As the scalaron develops a non-vanishing velocity during inflation, such a chiral coupling induces an effective chiral chemical potential during inflation and reheating, which can lead to the asymmetric production of heavy singlet fermions~\cite{Adshead:2018oaa}.
Moreover, since the chiral fermions in the Standard Model are charged under the SM gauge group and gravity, its chiral current coupling implies the Chern--Simons coupling of the scalaron via the Adler--Bell--Jackiw anomaly~\cite{Domcke:2018eki}.
Hence, the nonzero scalaron velocity would induce the instability to one polarization of the SM gauge fields~\cite{Turner:1987bw,Garretson:1992vt,Anber:2006xt} or the graviton~\cite{Lue:1998mq}, which opens up the various phenomenological implications such as enhanced scalar/tensor perturbations~\cite{Cook:2011hg,Barnaby:2011qe}, magnetogenesis~\cite{Turner:1987bw,Garretson:1992vt,Anber:2006xt}, chiral gravitational wave~\cite{Cook:2011hg,Barnaby:2011qe,Barnaby:2011vw,Anber:2012du,Domcke:2016bkh}, baryogenesis~\cite{Anber:2015yca,Fujita:2016igl,Kamada:2016eeb,Jimenez:2017cdr,Domcke:2019mnd,Domcke:2022kfs}, gravi-leptogenesis~\cite{Alexander:2004us}, to name a few.
We leave the detailed study of these effects for future work.

\section*{Acknowledgement}
M.\,He was supported by IBS under the project code, IBS-R018-D1.
M.\,Hong was supported by Grant-in-Aid for JSPS Fellows 23KJ0697.
K.\,M.\, was supported by JSPS KAKENHI Grant No.\ JP22K14044.

\appendix

\section{Convention}
\label{app-convention}

We adopt the sign convention $ (-1,1,1,1) $ for the spacetime metric $ g_{\mu\nu} $.
The Levi-Civita connection associated with $ g_{\mu\nu} $ is uniquely determined by imposing the metricity and torsionless conditions as in Eq.~\eqref{eq-levi-civita-connection}
\begin{align}\label{eq-levi-civita-connection}
  \Gamma\indices{^\rho_{\mu\nu}} \equiv \frac{1}{2} g^{\rho \sigma} \left( \partial_{\mu} g_{\nu\sigma} + \partial_{\nu} g_{\sigma \mu} - \partial_{\sigma} g_{\mu\nu}  \right) ~.
\end{align}
The covariant derivative associated with a general affine connection $ \bar{\Gamma} $ is defined as
\begin{align}
  \bar{\nabla}_{\mu} A\indices{^\nu_\rho} = \partial_{\mu} A\indices{^\nu_\rho} + \bar{\Gamma}\indices{^\nu_{\mu\lambda}} A\indices{^\lambda_\rho} - \bar{\Gamma}\indices{^\lambda_{\mu\rho}} A\indices{^\nu_\lambda}~.
\end{align}
The Riemann tensor, Ricci tensor, and Ricci scalar determined by $ \bar{\Gamma}\indices{^\rho_{\mu\nu}} $ are, respectively,
\begin{align}
  \bar{R}\indices{^\rho_{\sigma\mu\nu}} (\bar{\Gamma}) &\equiv \partial_{\mu} \bar{\Gamma}\indices{^\rho_{\nu\sigma}} - \partial_{\nu} \bar{\Gamma}\indices{^\rho_{\mu\sigma}} + \bar{\Gamma}\indices{^\rho_{\mu\lambda}} \bar{\Gamma}\indices{^\lambda_{\nu\sigma}} - \bar{\Gamma}\indices{^\rho_{\nu\lambda}} \bar{\Gamma}\indices{^\lambda_{\mu\sigma}} ~,\nonumber\\
  \bar{R}_{\mu\nu} (\bar{\Gamma}) &\equiv \bar{R}\indices{^\rho_{\mu\rho\nu}} (\bar{\Gamma}) ~,\nonumber\\
  \bar{R} (\bar{\Gamma}) &\equiv g^{\mu\nu} \bar{R}_{\mu\nu} (\bar{\Gamma}) ~.
\end{align}
Torsion is defined as the antisymmetric part of the connection as in Eq.~\eqref{eq-torsion} which we repeat here
\begin{equation}
  T\indices{^\rho_{\mu\nu}} \equiv \bar{\Gamma}\indices{^\rho_{\mu\nu}} - \bar{\Gamma}\indices{^\rho_{\nu\mu}} =C\indices{^\rho_{\mu\nu}} -C\indices{^\rho_{\nu\mu}} = - T\indices{^\rho_{\nu\mu}} ~.
\end{equation}
The torsion tensor can be decomposed in to three independent parts as
\begin{alignat}{2}
  &\text{(vector)}& \quad
  T_{\mu} &\equiv T\indices{^\alpha_{\mu\alpha}}
  ~,\nonumber\\
  &\text{(axial vector)}& \quad
  S^{\beta} &\equiv E^{\mu\nu\alpha\beta} T_{\mu\nu\alpha}
  ~,\nonumber\\
  &\text{(tensor)}& \quad
  q_{\alpha\beta\gamma} & \equiv T_{\alpha\beta\gamma} - \frac{1}{3} \left( g_{\alpha\gamma} T_{\beta}- g_{\alpha\beta} T_{\gamma} \right) +\frac{1}{6} E_{\alpha\beta\gamma\mu} S^{\mu} ~.
  \label{eq-decomposition-T}
\end{alignat}
where $ E^{\mu\nu\rho\sigma} $ is the totally anti-symmetric tensor with $E^{\mu\nu\rho\sigma} = \epsilon^{\mu\nu\rho\sigma} / \sqrt{-g}$, $E_{\mu\nu\rho\sigma} = - \sqrt{-g}\epsilon_{\mu\nu\rho\sigma} $, and $\epsilon^{0123} = 1$.

\section{Conformal mode}
\label{app-conformal}
In this appendix, we show the extraction of conformal mode of the spacetime metric~\cite{Ema:2020zvg} in the presence of the N--Y and the Holst terms, which can be useful when involving frame transformation and showing the appearance of the scalaron.

Given the spacetime metric in a specific frame denoted by ``$ \bullet $'', $ g_{\bullet \mu\nu} $, one can extract the conformal mode in this frame as
\begin{align}
  g_{\bullet \mu\nu} = \frac{\Phi_\bullet^2}{6 \Mpl^2} \tilde{g}_{\mu\nu} ~,
\end{align}
where $ \Phi_\bullet $ is the conformal mode in the corresponding frame and $ \tilde{g}=\det \tilde{g}_{\mu\nu} =-1 $.
As a result, the Ricci scalar $ R_\bullet (g_\bullet) $ can be rewritten as
\begin{align}
  R_\bullet = \frac{6\Mpl^2}{\Phi_\bullet^2} \left( \tilde{R}(\tilde{g}) - \frac{6}{\Phi_\bullet} \tilde{\Box} \Phi_\bullet \right) ~.
\end{align}
Besides, the components of torsion tensor should change accordingly.
The relevant quantities are
\begin{align}
  \nabla_\mu T^\mu &= \frac{6\Mpl^2}{\Phi_\bullet^2} \tilde{g}^{\mu\nu} \left( \tilde{\nabla}_\mu \tilde T_\nu + \tilde T_\nu \partial_\mu \!\ln \frac{\Phi_\bullet^2}{6 \Mpl^2} \right) ~, \\
  S^\mu &= \frac{6\Mpl^2}{\Phi_\bullet^2} \tilde{S}^\mu \equiv \frac{6\Mpl^2}{\Phi_\bullet^2} \frac{\epsilon^{\nu\rho\sigma\mu}}{\sqrt{-\tilde{g}}} \tilde{g}_{\nu\alpha} \tilde T\indices{^\alpha_{\rho\sigma}} ~, \\
  \nabla_\mu S^\mu &= \frac{6\Mpl^2}{\Phi_\bullet^2} \left( \tilde{\nabla}_\mu \tilde{S}^\mu + \tilde{S}^\mu \partial_\mu \!\ln \frac{\Phi_\bullet^2}{6 \Mpl^2} \right) ~,
\end{align}
where $ \tilde{T}\indices{^\mu_{\nu\rho}} = T\indices{^\mu_{\nu\rho}} $ because torsion is defined solely by the affine connection, and $ \tilde{T}_\mu = T_\mu $ for the same reason.
The tilde for them is only for the uniformity of notation.
The indices of all the quantities with tilde are raised (lowered) by $ \tilde{g}^{\mu\nu} $ ($ \tilde{g}_{\mu\nu} $).
With these results, one can reexpress the actions investigated in this paper with conformal mode and analyze the dynamical degrees of freedom.

Firstly, let us take a look at the simplest case~\eqref{eq-action-fR}. After the Legendre transformation~\eqref{eq-fR-Legendre} and decomposition of $ \bar{R} $, one extract the conformal mode which leads to the following form
\begin{align}
  \begin{split}
  S= \int  \dd^4 x\, &\left[ \frac{\Phi_{\rm J}^2}{12} \frac{2f'}{\Mpl^2} \tilde R - \frac{\Phi_{\rm J}}{2} \frac{2f'}{\Mpl^2} \tilde \Box \Phi_{\rm J} + \frac{\Phi_{\rm J}^2}{6} \frac{2f'}{\Mpl^2} \left( \tilde \nabla_\mu \tilde T^\mu +\tilde T^\mu \partial_\mu \ln \frac{\Phi_{\rm J}^2}{\Mpl^2} \right) \right. \\
  &\left. -\frac{\Phi_{\rm J}^2}{18} \frac{2f'}{\Mpl^2} \left( \tilde{T}^2 - \frac{1}{16} \tilde{S}^2 \right) -\left( \frac{\Phi_{\rm J}^2}{6 \Mpl^2} \right)^2 \left( \chi f' -f \right) \right] ~,
  \end{split}
\end{align}
where we require $ f' > 0 $ for healthy sign of the graviton kinetic term and, again, the contribution from $ q_{\mu\nu\rho} $ is trivial so we have omitted it.
One can immediately see that the solution for $ \tilde{S}^\mu $ is trivial so we simply neglect it in the following.
Solving the constraints for $ \tilde T_\mu $, one can obtain
\begin{align}
  S= \int \dd^4 x\, \left[ \frac{\Phi_{\rm J}^2}{12} \frac{2f'}{\Mpl^2} \tilde R +\frac{1}{2} \tilde{g}^{\mu\nu} \partial_\mu \sqrt{\frac{2f'}{\Mpl^2} \Phi_{\rm J}^2} \partial_\nu \sqrt{\frac{2f'}{\Mpl^2} \Phi_{\rm J}^2} - \left( \frac{2f'}{\Mpl^2} \frac{\Phi_{\rm J}^2}{6 \Mpl^2} \right)^2 \Mpl^4 \left( \frac{\chi}{4f'} - \frac{f}{4f'^2} \right) \right] ~,
\end{align}
from which one can immediately notice the fact that there is no additional kinetic terms for fields other than the gravitons and the new conformal mode $\Phi_\text{E} \equiv  2f' \Phi_{\rm J}^2 /\Mpl^2 $.
Therefore, one can conclude that this theory contains no dynamical scalaron.
More explicitly, one may calculate the rank of the kinetic matrix of the scalar sector in this theory including the conformal mode, which is trivial because the target space is one-dimension and full rank.
To see it in a more familiar language, one can absorb the new conformal mode to obtain
\begin{align}
  S= \int \sqrt{-g'} \dd^4 x\, \left[ \frac{\Mpl^2}{2} R' (g') -\Mpl^4 \left( \frac{\chi}{4f'} - \frac{f}{4f'^2} \right) \right] ~,
\end{align}
where we have defined
\begin{align}
  g'_{\mu\nu} \equiv
  \frac{\Phi_\text{E}^2}{6 \Mpl^2}\tilde{g}_{\mu\nu} ~,
\end{align}
which, as one may notice, is nothing but the Einstein frame metric.
In other words, such a theory is equivalent to an Einstein--Hilbert action plus a scalar potential without any kinetic term for this scalar field, \textit{i.e.}, non-dynamical scalaron.
This discussion shows the power of the conformal mode that whether there are new degrees of freedom or not can be shown in a frame-independent way.

To see a non-trivial example, let us take the case~\eqref{eq-NY-Star-deform} where scalaron becomes dynamical as shown in the main text.
By extracting the conformal mode, one finds
\begin{align}
  S = \int  \dd^4 x\, \left[ \frac{\Omega^2 \Phi_{\rm J}^2}{12} \tilde R -\frac{1}{2} \tilde{g}^{\mu\nu} \left( -\partial_\mu \!\left(\Omega \Phi_{\rm J}\right) \partial_\nu \!\left(\Omega \Phi_{\rm J}\right) + \frac{\alpha^2}{4} (\Omega\Phi_{\rm J})^2 \partial_\mu \!\ln \Omega^2 \, \partial_\nu \!\ln \Omega^2 \right) - \left( \frac{\Omega^2 \Phi_{\rm J}^2}{6 \Mpl^2} \right)^2 \frac{\chi^2}{\alpha_{\rm R}\Omega^4} \right]~.
\end{align}
Similarly, regarding $ \Omega \Phi_{\rm J} $ as the new conformal mode, one can calculate the determinant of the kinetic matrix for the field space vector $ ( \Omega \Phi_{\rm J} , \ln \Omega^2) $ as
\begin{align}
  \det
  \begin{pmatrix}
    -1 & 0 \\
    0 & \frac{\alpha^2}{4} (\Omega \Phi_{\rm J})^2
  \end{pmatrix}
  = -\frac{\alpha^2}{4} (\Omega \Phi_{\rm J})^2 ~,
\end{align}
which is non-vanishing for $ \alpha \neq 0 $.
This means that a dynamical scalar degree of freedom, scalaron, comes into the picture, which is consistent with the discussion in the main text.
If $ \alpha \to 0 $ (more precisely $ \alpha \to (12\alpha_{\rm R})^{-1/2} $ for scalaron mass approaching Planck scale), the scalaron becomes so heavy that it decouples with the system, so we should integrate out the scalaron leaving just the Einstein--Hilbert action.

\section{Other limits of the general form with Nieh--Yan term}
\label{app-other}
In this appendix, we discuss the limits of one or two parameter(s) in (\ref{eq-NY_general}) being zero, by taking corresponding limits from (\ref{eq-pX}) and (\ref{eq-Omega_constraint}).

Taking $\alpha_{R} \to 0$ and $\alpha_{RS} \to 0$ in \eqref{eq-NY_general} means one only has the $(\nabla_\mu S^\mu)^2$ term as an additional term in the action.
This case can be examined by taking $\alpha_{R} \to 0$ in \eqref{eq-pX} and \eqref{eq-Omega_constraint}, which has been illustrated in the main context.
Another limit which can be easily achieved is $\alpha_\text{S} \to 0$.
One simply takes $\beta \to - \alpha_\text{R} \alpha^2 / 4$ in (\ref{eq-pX}) and (\ref{eq-Omega_constraint}), which is still a $p(\phi,X)$ theory.

A less obvious limit is taking $\alpha_\text{R} \to 0$ in (\ref{eq-NY_general}).
This means one must fix $\alpha_\text{RS} = \alpha \alpha_\text{R}$ while sending $\alpha_\text{R}$ to zero.
Recalling $\beta \equiv \alpha_\text{S} - \alpha^2_\text{RS} / (4 \alpha_\text{R})$ one obtains
\begin{equation}
  \Omega^2 \to
  \frac{- \partial_\mu \Sigma \partial^\mu \Sigma + \frac{\alpha_\text{S}}{2 \alpha_\text{RS}^2} \Mpl^4 + \sqrt{\frac{2}{3}} \frac{ \Mpl^3}{4 \alpha_\text{RS}} \Sigma}
  {\frac{\alpha_\text{S}}{2 \alpha_\text{RS}^2} \Mpl^4 + \sqrt{\frac{2}{3}} \frac{ \Mpl^3}{8 \alpha_\text{RS}} \Sigma}
\end{equation}
from (\ref{eq-Omega_constraint}) and the action (\ref{eq-pX}) becomes
\begin{equation}
  S_\text{E} = \int \sqrt{-g_\text{E}} \dd^4 x\, \left[
    \frac{\Mpl^2}{2} R_\text{E}
    - \frac{1}{2 \Omega^4} \partial_\mu \Sigma \partial^\mu \Sigma
    + \frac{\Mpl^4}{4} \frac{\alpha_\text{S}}{\alpha_\text{RS}^2} \qty(1 - \frac{1}{\Omega^2})^2
    + \sqrt{\frac{2}{3}} \frac{ \Mpl^3}{8 \alpha_\text{RS}} \Sigma \frac{1}{\Omega^2} (\frac{1}{\Omega^2} - 1)
    \right]~.
\end{equation}
To see the $\alpha_\text{R} \to 0$ and $\alpha_\text{S} \to 0$ limit, one simply eliminates the term containing $\alpha_\text{S}$ in the results above.

Now we only have two limits remained, \textit{i.e.}, the $\alpha_\text{RS} \to 0$ limit and the $\alpha_\text{RS} \to 0$ and $\alpha_\text{S} \to 0$ limit.
To examine the former, \textit{i.e.}, the limit of the action without the $\bar R \nabla_\mu S^\mu$ term, one can take $\alpha \to 0$ limit in (\ref{eq-pX}) and (\ref{eq-Omega_constraint}).
One obtains
\begin{equation}
  \Omega^2 \to
  \frac{\partial_\mu \Sigma \partial^\mu \Sigma + \frac{\Mpl^4}{8 \alpha_\text{R}} + \frac{\Mpl^4}{48 \beta} \frac{\Sigma^2}{\Mpl^2}}
  {\frac{\Mpl^4}{8 \alpha_\text{R}}}
  ~,
\end{equation}
and the action becomes
\begin{equation}
  S_\text{E} = \int \sqrt{-g_\text{E}} \dd^4 x\, \left[
    \frac{\Mpl^2}{2} R_\text{E}
    - \frac{1}{2 \Omega^4} \partial_\mu \Sigma \partial^\mu \Sigma
    - \frac{\Mpl^4}{16 \alpha_\text{R}} \qty(1 - \frac{1}{\Omega^2})^2
    - \frac{\Mpl^2}{96 \beta} \frac{\Sigma^2}{\Omega^4}
    \right]~,
\end{equation}
which is a $p(\phi,X)$-type theory.

To take the $\alpha_\text{RS} \to 0$ and $\alpha_\text{S} \to 0$ limit, one first set $\beta = - \alpha_\text{R} \alpha^2 / 4$ and send $\alpha$ to zero.
One can check the propagating mode disappears as discussed in Sec.~\ref{sec:absence}.


\bibliographystyle{utphys}
\bibliography{ref}

\end{document}